\documentclass[twocolumn,twoside,slac_two]{revtex4}
\usepackage{graphicx}
\usepackage{fancyhdr}
\begin{document}

\title{A Warning on the GeV-TeV Connection in Blazars}

%

\author{L. Costamante}
\affiliation{HEPL/KIPAC, Stanford University, Stanford, CA 94024, USA \\
Now at: Dept. of Physics, Universit\'a degli Studi di Perugia, I-06123 Perugia, Italy}

\begin{abstract}
Fermi-LAT spectra at high energies (HE, 0.1-100 GeV) are often extrapolated to very high energies 
(VHE, $\gtrsim$100 GeV),  and considered either a good estimate or an upper limit for the blazars 
intrinsic VHE spectrum.  This assumption seems not well justified, neither theoretically nor 
observationally. Besides being often softer, observations do indicate that spectra 
at VHE could be also harder than at HE, even when adopting the limit of $\Gamma\geq1.5$.
Results based on such straightforward GeV-TeV extrapolations are in general not reliable,
and should be considered with caution. 
\end{abstract}

\maketitle

\thispagestyle{fancy}

\section{Introduction}
The well-determined Fermi-LAT spectra (or Upper Limits, UL) 
in the MeV-GeV band for several TeV blazars are often used
to derive stronger constraints on the diffuse  extragalactic background light (EBL), 
or to constrain the distance for BL Lacs of uncertain redshift
(e.g. \cite{lat1553,ver1424,persic,orr,prandini}).

The underlying assumption of these studies is that the
extrapolation of the Fermi-LAT spectrum to the VHE band is
either a good estimate or an upper limit for the intrinsic VHE
spectrum of the source, since they belong to the same hump in the spectral 
energy distribution (SED).  The extrapolation is done either as a simple
extension of the power-law (or best fit) model for the Fermi-LAT data, 
or using a one-zone synchrotron self-Compton model (SSC) on the overall SED.

However, the observational evidence shows that this
assumption is not well justified anymore.

\section{Blazars gamma-ray spectra}
Blazars display a wide range in SED peak energies, as well as in emission components 
(both in time and space along the jet). The VHE band samples typically the
highest energies of the emitting particle distribution, and is therefore more sensitive 
to even small changes in the acceleration and cooling processes.
Observationally, irrespective of EBL absorption (i.e. adopting the same EBL model for all sources),
the band between HE and VHE is the energy range 
where the spectrum of blazars typically changes the most, either because of the closeness of the SED peak
or of the end of the emitting particles distribution.
Therefore, the extrapolation to VHE of the HE slope is in general never a good assumption. 

This is particularly true for the high-energy-peaked BL Lacs (HBL) which are bright in Fermi
and have been easily detected in the first years of operation \cite{sanchez,brightAGN,1fgl}.
These are characterized by the high-energy SED peak being very close to the LAT band
(e.g. Mkn\,421, Mkn\,501, PKS\,2155-304, PKS\,2005-489, 1ES\,1553+113 etc).
For them, the HE and VHE bands sample the two sides of the gamma-ray hump
in the SED, and thus the intrinsic VHE spectra are much softer than the HE spectra
Indeed this is the most common case among Fermi-detected HBL \citep{2lac,2fgl}.

The main question however is: can the VHE spectrum be harder than the HE spectrum ?  
Can thus the Fermi-LAT index be used as reasonable upper limit for the hardness 
of the VHE spectrum, or for its luminosity ?  
Concave HE-VHE spectra have not been observed (yet) directly.
However, the observational (as well as theoretical) evidence is now
showing this to be possible.

\begin{figure*}
\includegraphics[width=75mm]{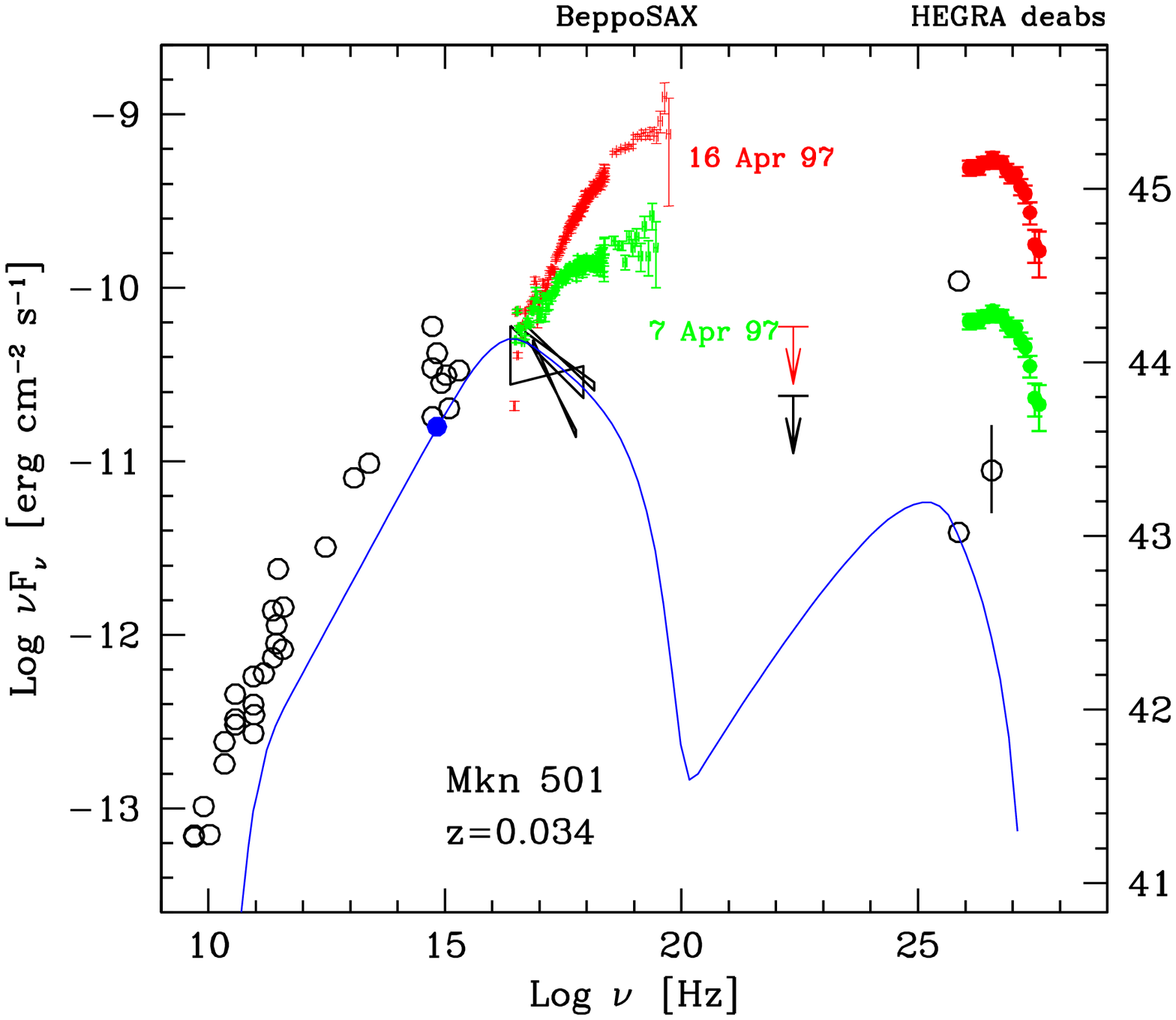}
\includegraphics[width=80mm]{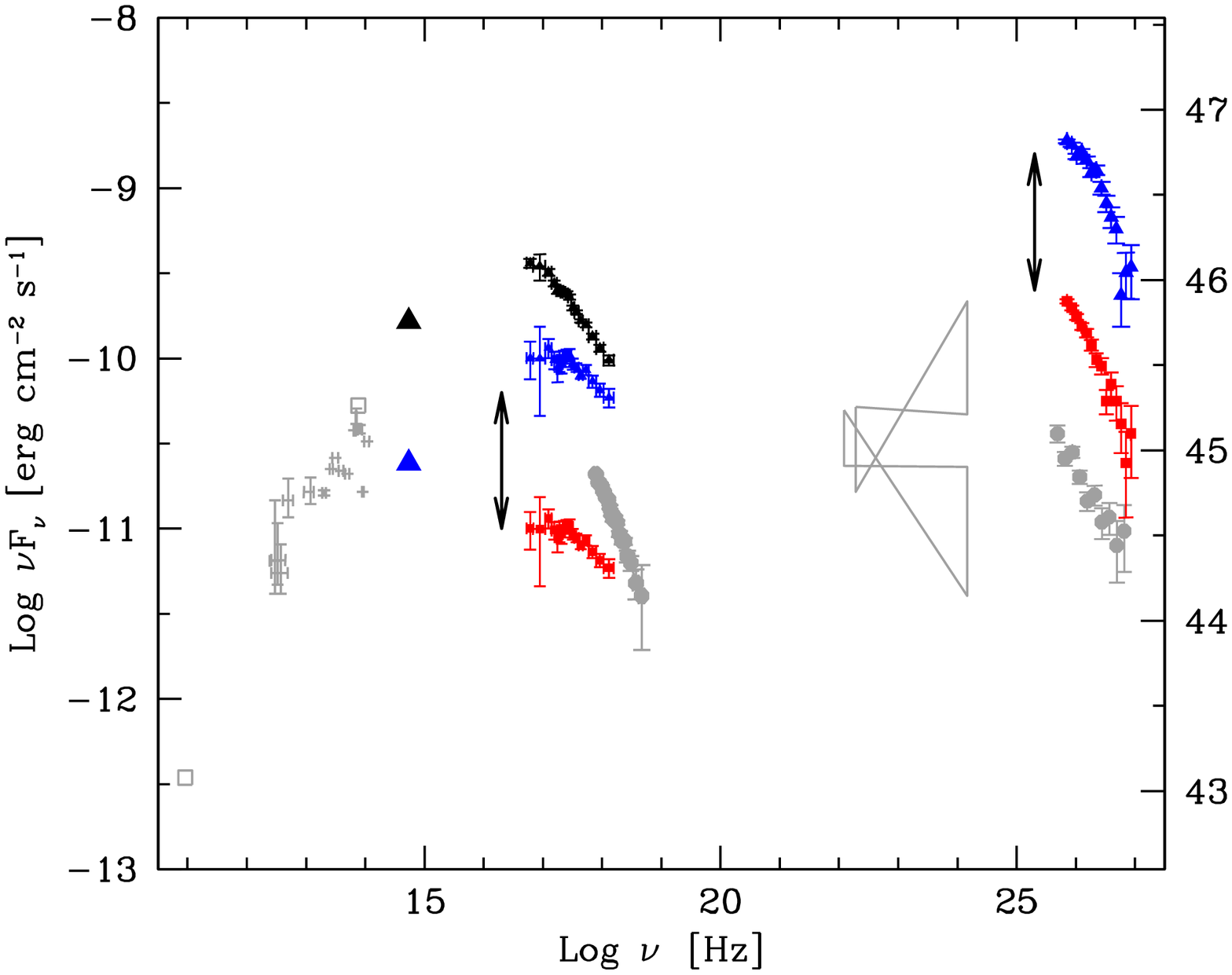}
\caption{Examples of new emission components in the SED of blazars
emerging over previous (more steadier) SEDs.
Left: Mkn\,501 in 1997 \cite{pian,hegra501}.
Right: PKS\,2155-304 in 2006 \cite{chandra2155}.
The peculiar spectral and timing properties of this 2006
VHE flare was interpreted as superposition of two SEDs from two different zones. 
The observed SED cannot be explained with a single SSC component (see \cite{chandra2155}).
}
\label{f1}
\end{figure*}

\section{TeV spectra can be harder than GeV spectra}
\begin{figure*}
\includegraphics[width=80mm]{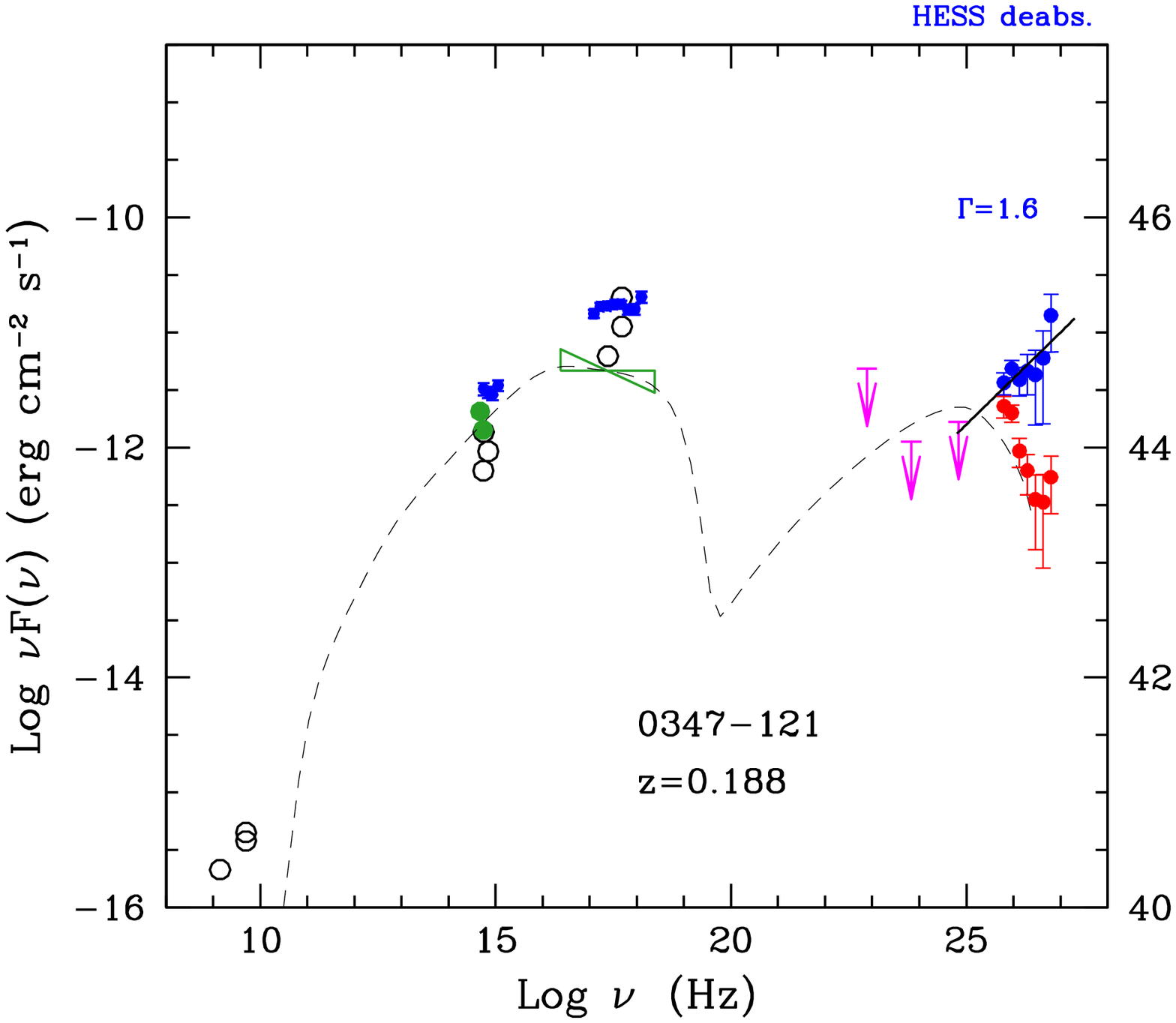}
\includegraphics[width=80mm]{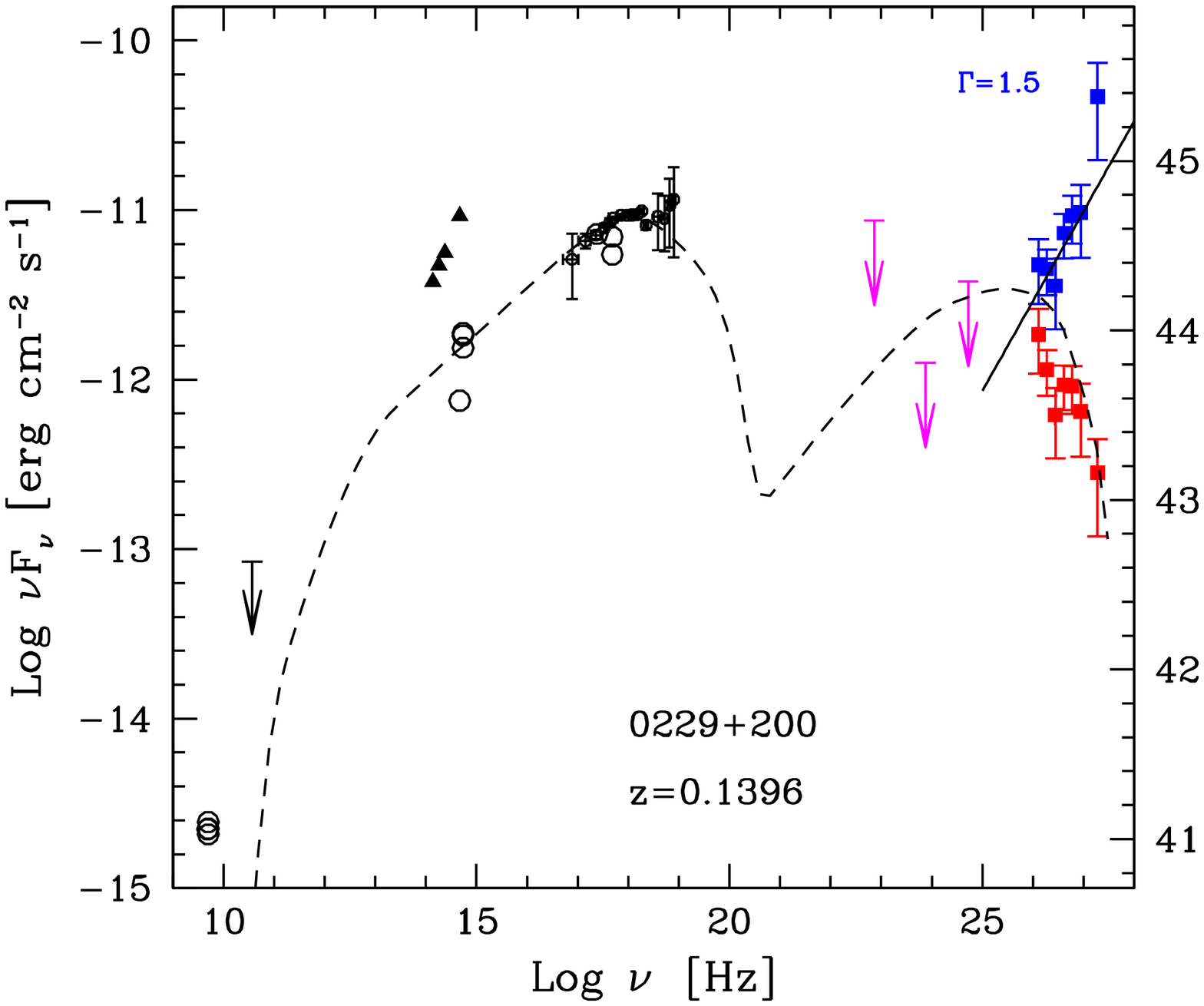}
\caption{Examples of SED of selected HBL, characterized by hard TeV spectra with $\Gamma\sim1.5-1.6$
irrespective of the level of EBL.
The absorption-corrected VHE spectra (blue points) shown here are corrected using the EBL calculations
by \cite{franceschini} (i.e. a low EBL). Very similar results are obtained using calculations by 
\cite{dominguez,gilmore}. Higher levels of the EBL would make the intrinsic spectra
even harder.  Data and historical modeling from 
\cite{extreme,sax0229,1fgl,0229,0347,gcc}.  }
\label{f3}
\end{figure*}

Let us consider three facts. 
\begin{enumerate}

\item BL Lacs do show multiple spectral components in their synchrotron emission,
which traces directly the shape of the electron distribution(s). 
In several cases we have already seen the superposition of two different emission components 
at high electron energies, with a new component emerging over a previous/steadier SED.
Classical examples are given by Mkn 501 in 1997  \citep{pian,hegra501},
1ES\,1959+650 in 2002 \cite{orphan} and PKS\,2155-304 in 2006 \cite{chandra2155} 
(see e.g. Fig. \ref{f1}).
The same can thus happen in the inverse Compton emission.

\item The superposition of multiple spectral components is seen also outside specific flaring episodes, 
on long (months to years) timescales.
One of the most evident cases was given by PKS\,2005-489 
in the synchrotron emission (Fig. \ref{f2}), during multi-wavelength campaigns in 2004-2005.
The X-ray band in 2005 became more and more dominated by a new harder emission component 
emerging in the SED, which probably reached its maximum in June 2009 \cite{2005mwl2}. 

\begin{figure}
\includegraphics[width=75mm]{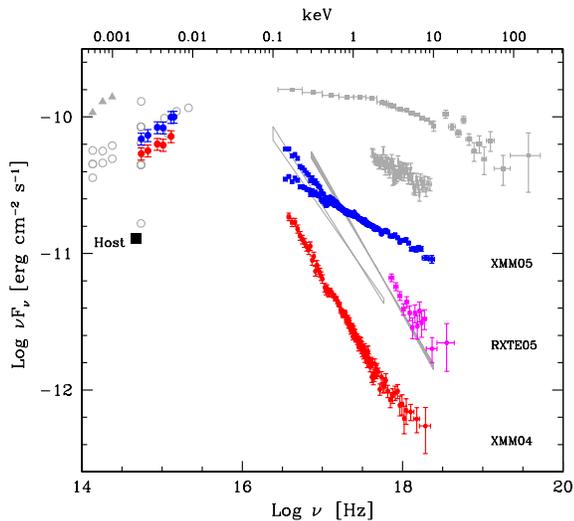}
\caption{SED of PKS\,2005-489 zoomed in the optical-X-ray band, with the opt-to X-ray 
data from different epochs.   Blue data correspond to 
the XMM spectrum in 2005 corrected for two different estimates of the Galactic 
N$_{\rm H}$ (5.08 or 3.93 $\times10^{20}$ cm$^{-2}$). 
Even with the lowest Galactic column density, in 2005 the X-ray spectrum cannot be joined 
to the UV data from the Optical Monitor with a single smooth spectral component \cite{hess2005}.
}
\label{f2}
\end{figure}

\item At VHE, intrinsic spectra as hard as $\Gamma=1.5-1.6$ are already observed, 
with a low EBL density  (even harder in case of higher EBL densities).
Classic examples are given by 1ES\,1101-232 \cite{nature,1101mwl}, 
1ES\,0347-121 \cite{0347} and especially 1ES\,0229+200 (Fig. \ref{f3}),
which is characterized by such hard VHE spectrum up to $\sim$10 TeV \cite{0229}.
This demonstrates that there exist  physical conditions in blazars which can yield 
TeV spectra  as hard as $\Gamma=1.5$, and with higher luminosity than in the synchrotron emission.
Such conditions can in principle form also in specific zones of the jet, and/or in specific epochs.
The overall SED of such components can easily remain hidden below a more "standard" SED 
and emerge or become dominant only at VHE.

\end{enumerate}

Considering these three facts together, it becomes clear that spectra at VHE
can very well be harder than at HE.  This is true especially for those HBL where the 
Fermi-LAT index is closer to 2 than 1.5 (or even steeper than 2, as in IBL and LBL).

Over such a wide range of energies (5 orders of magnitude), 
it seems  not only possible but even likely  that  a combination of different spectral 
components --either in time or from different particle populations or different emission mechanisms 
for the same particles-- can result in {\it concave} overall spectra. 
It seems only a matter of time (and statistics) before up-turns somewhere 
in the overall 100 MeV -- 10 TeV band are directly and significantly detected 
in the simultaneous gamma-ray spectrum of some blazars.

\begin{figure}
\includegraphics[width=80mm]{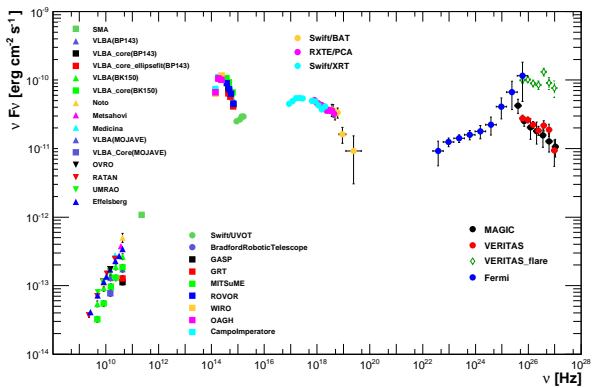}
\caption{SED of Mkn\,501 averaged over all observations taken during a multiwavelength 
campaign between March and August 2009 (from \cite{david501}). 
VHE data are corrected for EBL absorption according to \cite{franceschini}. 
The VERITAS data from a 3-day TeV flare are depicted separately in the plot.
}
\label{f4}
\end{figure}

It should not be surprising, because the observational ingredients are all there.
In fact, a possible example  is already given by the Fermi-LAT spectrum 
of Mkn\,501 \citep{david501},  where a flaring episode has already produced a time-average  HE spectrum 
apparently hardening towards higher energies (see Fig. \ref{f4}).

Therefore, in general, Fermi-LAT spectra cannot be reliably used as UL to 
a) derive constraints on redshift, or  b) put stronger limits on the intensity of the EBL. 
The results would be as weak/unreliable as their assumptions, and lead to gross over/under-estimates.

\section{Consequences for EBL and blazar studies}
For EBL and resdhift studies, the $\Gamma=1.5$ limit for the intrinsic photon 
index is still a more robust benchmark than any Fermi-LAT extrapolation.
Although it is not a ``hard" limit (theoretically there are many mechanisms to obtain harder spectra, 
see e.g. \cite{lefa}),  at present it still represents the borderline between 
reality and speculation (for a full discussion, see e.g. \cite{preston}). 

Observationally, intrinsic spectra with $\Gamma<1.5$ ($\pm0.2$) have never been observed so far
in blazars at high electron energies (e.g. $\gamma>10^{3-4}$), neither in the synchrotron emission
nor in the inverse Compton spectrum of low-redshift sources. 
Photon indexes of 1.2-1 are observed at X-ray energies in high-redshift, low-energy-peaked 
blazars (typically FSRQ), but these appear to be explained 
as low-energy cutoffs in the electron distribution, if not by internal absorption 
\cite{tavecchio,worsley,ggpowerful}.
In fact, this is a spectral feature that could in principle appear in the Fermi-LAT band 
for some TeV-peaked sources (i.e. those where the LAT band is deep in the valley between 
the two SED humps).   However, it would not automatically imply a similarly hard spectrum at VHE, 
because  in such case the VHE band would correspond to the particle distribution 
well above the low-energy cutoff.

At present the EBL spectrum between 0.1 and 10 $\mu$m is constrained rather well
(mostly within $\sim$50\%; \cite{nature,0229,latebl}), simply by using the 
range of spectra observed in blazars.
Any further improvement beyond that requires a prediction of the blazars' high-energy emission 
at a level of accuracy which seems not (yet) at hand \cite{preston}.
The uncertainty is  systematical, on which model and physical conditions
are actually working in blazars.  
For a given HE spectrum and overall SED,  the range of possible VHE spectra 
is still large ($\Delta\Gamma\approx2$), even in a SSC scenario,
depending on the choice of  parameters, zones and adopted  particle distributions  (see e.g. \cite{lefa}).
And observations are demonstrating that we are still missing some fundamental aspects 
of the blazar physics. 
One-zone SSC models can work {\it a-posteriori}, but cannot be used reliably {\it a-priori}
since blazars have multiple emission components, whose behaviour and interplay is still unknown.
A reliable prediction of the VHE spectrum from the HE one is therefore not yet possible,
at least in individual sources.  


For a real progress in this field, it seems now more useful and fruitful 
to fix the EBL to the most consistent and likely values (e.g. \cite{franceschini,dominguez,gilmore}),
and to focus on improving our understanding of the blazar emission properties.

\bigskip 

\end{document}